\documentclass[10pt,twocolumn,showpacs,preprintnumbers,amsmath,amssymb,aps,prl,longbibliography,superscriptaddress]{revtex4-2}
\usepackage{amsmath}
\usepackage{graphicx}
\usepackage{wasysym}
\usepackage{amssymb}
\usepackage{amsfonts}
\usepackage{bm}
\usepackage{natbib}
\usepackage{enumerate}
\usepackage{color}
\usepackage{hyperref}
\usepackage{multirow}

\newcommand{\beq}{\begin{equation}}
\newcommand{\eeq}{\end{equation}}
\newcommand{\bea}{\begin{eqnarray}}
\newcommand{\eea}{\end{eqnarray}}
\newcommand{\bes}{\begin{split}}
\newcommand{\ees}{\end{split}}

\usepackage{cleveref}
\crefname{appendix}{App.}{Apps.}
\crefname{equation}{Eq.}{Eqs.}
\crefname{figure}{Fig.}{Figs.}
\crefname{table}{Tab.}{Tabs.}
\crefname{section}{Sec.}{Secs.}

\mathchardef\nss="711B

\def\nss{\mathcal{S}}

\def\be{\begin{eqnarray}}
\def\ee{\end{eqnarray}}
\def\beq{\begin{equation}}
\def\eeq{\end{equation}}

\begin{document}
\title{Resistance Distribution of Decoherent Quantum Hall-Superconductor Edges}
\author{Yichen Hu}
\thanks{huy@fau.edu}
\affiliation{Department of Physics, Princeton University, Princeton, New Jersey 08544, USA}
\affiliation{Department of Physics, Florida Atlantic University, Boca Raton, Florida 33433,USA}

\author{Jing Wang}
\thanks{wjingphys@fudan.edu.cn}
\affiliation{State Key Laboratory of Surface Physics and Department of Physics, Fudan University, Shanghai 200433, China}
\author{Biao Lian}
\thanks{biao@princeton.edu}
\affiliation{Department of Physics, Princeton University, Princeton, New Jersey 08544, USA}

\begin{abstract}
We study the probability distribution of the resistance, or equivalently the charge transmission, of a decoherent quantum Hall-superconductor edge, with the decoherence coming from metallic puddles along the edge. Such metallic puddles may originate from magnetic vortex cores or other superconductivity suppressing perturbations. In contrast to the distribution of a coherent edge which is peaked away from zero charge transmission, we show analytically and numerically that the distribution of a decoherent edge with metallic puddles is always peaked at zero charge transmission, which serves as a probe of coherence of superconducting chiral edge states. We further show that the distribution width decays exponentially in magnetic field and temperature. Our theoretical decoherent distribution agrees well with the recent experimental observation in graphene with superconducting proximity.
\end{abstract}

\maketitle

The superconducting proximity of quantum Hall (QH) or quantum anomalous Hall states \cite{Chang_2013,Checkelsky_2014,kou2014,mogi2015,deng2020} has been attracting growing interests of study, for their promise of realizing topological superconductors and other novel topological states, and their potential applications in quantum information \cite{Mong2014,Clarke2014,Clarke2013,Hu2018,Teo2023,Lian2018}. Recent experiments \cite{Rickhaus2012,Komatsu2012,Wan2015,Amet2016,Lee2017,Park2017,Sahu2018,Matsuo2018,Kozuka2018,Seredinski2019,Zhao2020,Zhao2023,Yuval2022,Hatefipour2022,Uday2023} have made prominent efforts towards revealing the ubiquitous transport signatures of superconducting edge states. A significant goal is to achieve coherent chiral Bogoliubov (or Majorana) edge modes, which would allow the coherent Andreev interference in the resistance along an edge between QH and superconductor (SC) \cite{Lian2016,Wangj2018,Lian2019,Hoppe2000,Giazotto2005,Akhmerov2007,vanOstaay2011}. Moreover, quantized transport and interference signatures can be achieved for the single chiral Majorana edge mode of $p+ip$ topological SC \cite{Fu2009,Akhmerov2009,Qi2010,Wang2015,Chung2011,Lian20182}. While coherent interferences have been observed in QH states without superconductivity in GaAs \cite{Nakamura2019} and graphene \cite{Deprez2021,Ronen2021}, the experimental signature of resistance interference of chiral QH-SC edges (in graphene) remains illusive \cite{Zhao2020,Zhao2023}, due to various complications such as effectiveness of proximity, magnetic vortices, thermal fluctuations, etc \cite{Manesco_2022,kurilovich2201,zhao2024}.

In this Letter, we study the effect of possible random metallic puddles along the superconducting edge of a QH-SC-QH junction, which can be the leading factor of decoherence at low temperatures. We show that the resistance of such a system is determined by the generalized Landauer-B\"uttiker formula, in which the random metallic puddles play the role of floating leads. In the absence of metallic puddles, regarding the coherent interference phase angle as random (due to elastic disorders and variation of system parameters), the charge transmission $t_\text{eh}$ (defined in \cref{eq:teh}) shows a probability distribution peaked at its maximum and minimum. In contrast, with decoherent metallic puddles, the charge transmission probability distribution will be solely peaked at zero, which is consistent with the experiment \cite{Zhao2023}. Assuming the metallic puddles arise from magnetic vortex cores and other SC suppressing perturbations, we estimate the number of metallic puddles to be linear in magnetic field $B$ and temperature $T$, leading to a charge transmission suppressed exponentially in $B$ and $T$, in agreement with the experiment \cite{Zhao2023}.

\begin{figure}[tbp]
\centering
\includegraphics[width=3.4in]{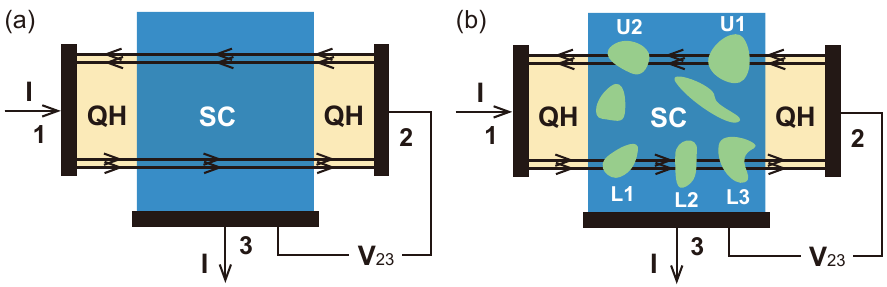}
\caption{The QH-SC-QH junction (a) without decoherence; and (b) with superconducting edge states encountering decoherent metallic puddles (green regions) in SC.}
\label{fig1}
\end{figure}

\emph{Coherent edge transport.} By stacking a normal SC onto the middle region of a 2D integer QH insulator with nonzero Chern number $C$, one can experimentally implement a QH-SC-QH heterojunction in \cref{fig1}(a) \cite{Zhao2020,Zhao2023}. We consider the middle region of the heterojunction to be in the same topological phase as the Chern number $C$ QH state with broken charge U(1) symmetry, namely, a superconductor with Bogoliubov-de Gennes (BdG) Chern number $C_\text{BdG}=2C$, which would be the case when the SC proximity gap is smaller than the insulating QH state bulk gap \cite{Qi2010,Wang2015}. Accordingly, the $C$ chiral electron edge modes of the QH state become $2C$ chiral Bogoliubov (or equivalently Majorana) edge modes.

Assume $c_j$ and $c_j^\dag$ ($1\le j\le C$) are the annihilation and creation operators of the $C$ chiral electron modes on the QH edge which has spatial coordinate $x$. In the absence of decoherence, these chiral edge states with SC proximity have an action \cite{Wangj2018,Lian2019}
\beq\label{eq:action}
S=\int dt dx \frac{1}{2}\Big[\sum_{j=1}^{2C} i \gamma_j (\partial_t+v_j\partial_x)\gamma_j+i\bm{\gamma}^T M(x)\bm{\gamma}\Big],
\eeq
where $\gamma_j$ ($1\le j\le 2C$) are the Majorana modes defined such that $c_j=(\gamma_{2j-1}+i\gamma_{2j})/\sqrt{2}$, which have velocities $v_j$. We organize them as a vector $\bm{\gamma}=(\gamma_1,\cdots,\gamma_{2C})^T$, and $M(x)$ is a $2C\times2C$ real anti-symmetric matrix with matrix elements given by the potentials and pairing amplitudes experienced by the edge modes \cite{lian_chiral_edge2022}, which can be $x$-dependent if there are static disorders. 

Generically, a chiral Bogoliubov fermion mode at zero energy with creation operator $f^\dag=\sum_{j=1}^{2C}\int dx \zeta_j(x)\gamma_j$ has amplitudes $\zeta_j$ satisfying $v_j\partial_x \zeta_j(x) + \sum_{j'=1}^{2C}M_{jj'}(x)\zeta_{j'}(x)=0$. 
Solving this equation yields spatial oscillations of $\zeta_j(x)$ with respect to $x$. For $M(x)$ independent of $x$ (disorderless), the spatial oscillation frequencies are simply given by the momentum differences between different Bogoliubov eigenmodes, which depends on chemical potential and pairing amplitudes, etc \cite{Lian2016,Wangj2018,Lian2019}. We denote the normal transmission and Andreev (electron-to-hole conversion) transmission coefficients of the edge as $t\in[0,C]$ and $t_A\in[0,C]$, respectively, which satisfy $t+t_A=C$ due to the absence of back-scattering. For later purposes, we define the charge transmission fraction $t_{\text{eh}}=\frac{t-t_A}{C}=1-\frac{2t_A}{C}\in[-1,1]$. The spatial oscillations of amplitudes $\zeta_j(x)$ then lead to $t_{\text{eh}}$ of the form
\begin{equation}\label{eq:teh}
t_{\text{eh}}=\frac{t-t_A}{C}\simeq A_0+\sum_{m}A_{m}\cos\phi_m\ ,
\end{equation}
where $A_0$ and $A_m$ are constants, and $\phi_m$ are a set of phase angles dependent (linearly to the lowest order) on the edge length $L_\text{edge}$ and gate voltage $V_G$ (chemical potential) on the edge, given other parameters (pairing, static disorders, etc) fixed. 

Specifically, for the experiments \cite{Zhao2020,Zhao2023} which utilize the Chern number $C=2$ state of graphene, if the velocities $v_j$ in \cref{eq:action} are approximately equal, $t_{\text{eh}}$ would approximately only depend on one phase angle $\phi$ (see supplementary material (SM) \cite{suppl}):
\begin{equation}\label{eq:teh-2}
t_{\text{eh}}\simeq A_0+A_1\cos\phi\ .
\end{equation}
This is because the amplitudes $\zeta_j(x)$ of the $2C=4$ chiral Majorana modes undergo a spatial SO(4)$\simeq$SU(2)$\otimes$SU(2) rotation \cite{Yang1990}: the first SU(2) rotates in the electron-hole space and contributes a phase angle $\phi$, while the second SU(2) generates charge conserving rotations among the two electron modes and has not effect on $t_{\text{eh}}$.

In the absence of decoherence (elastic scatterings from static disorders are allowed), the transport of the QH-SC-QH heterojunction in \cref{fig1}(a) is governed by the generalized Landauer-B\"uttiker formula \cite{blonder1982,anantram1996,entin2008}
\begin{equation}\label{eq:coherent-LB}
\begin{split}
I_1&=\frac{Ce^2}{h}\left[(V_1-V_\text{sc})+t_{\text{eh}}^{(u)}(V_\text{sc}-V_2)\right]\ ,\\
I_2&=\frac{Ce^2}{h}\left[(V_2-V_\text{sc})+t_{\text{eh}}^{(l)}(V_\text{sc}-V_1)\right]\ ,\\
I_{3}&=I_\text{sc}=-I_1-I_2\ ,
\end{split}
\end{equation}
where $h$ is the Planck constant, $e$ is the electrical charge, $t_{\text{eh}}^{(u)}$ and $t_{\text{eh}}^{(l)}$ are the charge transmission fractions (\cref{eq:teh}) of the upper and lower superconducting edges in \cref{fig1}(a), respectively. Leads $1$ and $2$ are connected to the edges of the left and right QH regions, respectively, and lead $3$ is connected to the SC bulk. $I_j$ and $V_j$ are the inflow current and voltage of lead $j$, and particularly $V_3=V_\text{sc}$ is the SC voltage. \cref{eq:coherent-LB} assumed zero normal and Andreev reflection coefficients $r$ and $r_A$ \cite{Lian2019}, due to the absence of back-scattering. Note that in experiment \cite{Zhao2020}, the upper edge has no SC proximity, which amounts setting $t_{\text{eh}}^{(u)}\equiv1$. In experiment \cite{Zhao2023}, both edges have SC proximity.

The experiments \cite{Zhao2020,Zhao2023} set $I_1=-I_3=I$ (thus $I_2=0$) and measured the non-local differential resistance $R_d=\frac{dV_{23}}{dI}$ with $V_{23}=V_2-V_\text{sc}$. In the linear response regime, solving \cref{eq:coherent-LB} gives 
\begin{equation}\label{eq:Rd}
R_d=\frac{V_2-V_\text{sc}}{I}=\frac{h}{C e^2}\frac{t_{\text{eh}}^{(l)}}{1-t_{\text{eh}}^{(u)}t_{\text{eh}}^{(l)}}\ .
\end{equation}
Following the experiment \cite{Zhao2020,Zhao2023}, we then define a quantity $\mathcal{P}_{\text{eh}}=\frac{R_d}{R_d+\frac{h}{Ce^2}}=\frac{t_{\text{eh}}^{(l)}}{1+t_{\text{eh}}^{(l)}-t_{\text{eh}}^{(u)}t_{\text{eh}}^{(l)}}$. The experiment showed that $|\mathcal{P}_{\text{eh}}|\lesssim 0.05$, and $\mathcal{P}_{\text{eh}}$ as a function of gate voltage $V_G$ obeys a probability distribution with a triangular shape peak at $\mathcal{P}_{\text{eh}}=0$ and no other peaks.

\begin{figure}[tbp]
\centering
\includegraphics[width=3.4in]{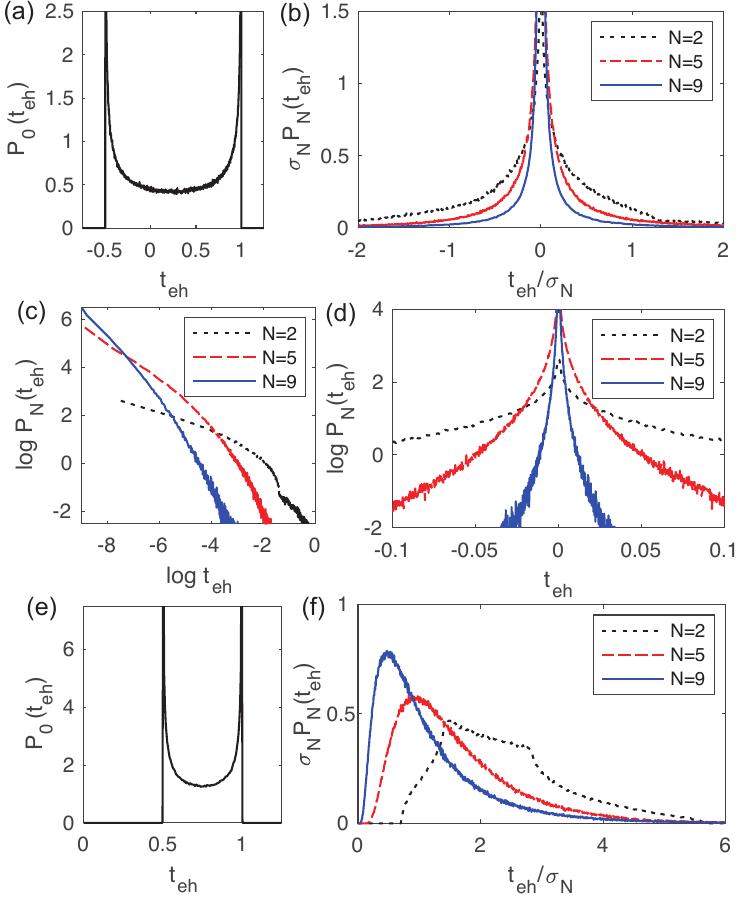}
\caption{ The Monte Carlo numerical results for (a)-(d) $A_0=0.25$, $A_1=0.75$, and (e)-(f) $A_0=0.75$, $A_1=0.25$. Panels (a) and (e) show $P_0(t_{\text{eh}})$, (b) and (f) show different $P_N(t_{\text{eh}})$, in which $\sigma_N$ stands for its standard deviation. (c) and (d) show the log-log and semi-log plots of panel (b), respectively. 
}
\label{fig2}
\end{figure}

We first examine if the measured $\mathcal{P}_{\text{eh}}$ is understandable by the Landauer-B\"uttiker framework without decoherence (i.e., \cref{eq:coherent-LB}). Since the measured $|\mathcal{P}_{\text{eh}}|\ll 1$ \cite{Zhao2020,Zhao2023}, this implies $|t_{\text{eh}}^{(l)}|\ll 1$ and therefore
\begin{equation}\label{eq:Peh}
\mathcal{P}_{\text{eh}}\approx t_{\text{eh}}^{(l)}\ .
\end{equation}
Thus, to the lowest order, $\mathcal{P}_{\text{eh}}$ obeys the same probability distribution as $t_{\text{eh}}^{(l)}$, which we investigate hereafter. Note that \cref{eq:Peh} becomes exact if $t_{\text{eh}}^{(u)}=1$, i.e. if the upper edge has no SC proximity, e.g. in experiment \cite{Zhao2020}. For $C=2$, assume $t_{\text{eh}}^{(l)}$ takes the form of \cref{eq:teh-2}, and assume the phase $\phi$ is uniformly randomly distributed in $[0,2\pi)$ as the gate voltage $V_G$ changes. This yields a probability distribution satisfied by $t_{\text{eh}}=t_{\text{eh}}^{(l)}$ (see SM \cite{suppl}):
\begin{equation}
P_0(t_{\text{eh}})=\begin{cases}
&\frac{1}{\pi \sqrt{A_1^2-(t_{\text{eh}}-A_0)^2}}\ , \ (t_{\text{eh}}\in [A_-,A_+])\\
&0\ ,\qquad \qquad\qquad\ \    (t_{\text{eh}}\notin [A_-,A_+])
\end{cases}
\label{P0}
\end{equation}
where $A_\pm=A_0\pm |A_1|$. Such a distribution is peaked at the maximal and minimal values $A_\pm$ of $t_{\text{eh}}$, as shown in \cref{fig2}(a),(e).
Even if one takes the most generic form of $t_{\text{eh}}$ in \cref{eq:teh}, assuming independently random phases $\phi_m\in[0,2\pi)$, one would end up with a probability distribution $P_0(t_{\text{eh}})$ similarly peaked at the maximum and minimum of $t_{\text{eh}}$, since the values of cosine functions $\cos\phi_m$ are the most probable at their maxima and minima.
This contradicts the experimental observation \cite{Zhao2023} that the probability distribution is solely sharply peaked at $t_{\text{eh}}=0$. Therefore, the experiment cannot be explained by coherent edge state transport.

\emph{Edge with decoherent metallic puddles}. Decoherence of the chiral edge states may originate from inelastic scatterings with other gapless degrees of freedom. If the surrounding QH and SC bulk gaps are nonzero, the only inelastic scatterings are from phonons, which will be suppressed to zero as the temperature $T\rightarrow 0$. However, small non-superconducting metallic puddles may (effectively) arise in the SC middle region as illustrated in \cref{fig1}(b) due to inhomogeneity, which would induce strong decoherence persistent to the zero temperature if encountered by the chiral edge states. A metallic puddle may be a place with ineffective SC proximity, or a magnetic vortex core if the QH state is realized in a magnetic field (as is true in the experiment \cite{Zhao2023}). Hereafter, we ignore phonons and consider only the decoherence from metallic puddles, as is legitimate at low temperatures.

Assume the lower (upper) edge encounters $N^{(l)}$ ($N^{(u)}$) metallic puddles as illustrated in \cref{fig1}(b). Each of these metallic puddles plays the role of a floating metallic lead. Together with the original leads $1$ and $2$, we have the chiral edge states connected to in total $N^{(l)}+N^{(u)}+2$ leads, which we relabel by an index $n=1,2,\cdots, N^{(l)}+N^{(u)}+2$ counterclockwisely, with $n=1$ identified with the original lead $1$ in \cref{fig1}(b). Thus, the generalized Landauer-B\"utikker formula becomes
\begin{equation}\label{eq:coherent-LB-N}
\begin{split}
I_{n}'&=\frac{Ce^2}{h}\left[(V_n'-V_\text{sc})+t_{\text{eh},n}(V_\text{sc}-V_{n-1}')\right]\ , \\
\end{split}
\end{equation}
where $I_{n}'$ and $V_n'$ are the inflow current and voltage of the \emph{newly} defined $n$-th lead, and $t_{\text{eh},n}$ is the charge transmission fraction of the coherent piece of edge between leads $n-1$ and $n$, with $n$ identified with $n+N^{(l)}+N^{(u)}+2$. The SC voltage is $V_\text{sc}$. The inflow current through the original lead $3$ to SC is still $I_3=I_\text{sc}=-\sum_{n=1}^{N^{(l)}+N^{(u)}+2}I_n'$ by charge conservation.

Physically, we should set inflow current $I_n'=0$ for $n\neq 1$ and $n\neq N^{(l)}+2$, which are the floating leads in the SC (metallic puddles) and thus disconnected from external current sources. \cref{eq:coherent-LB-N} then implies $\frac{V_n'-V_\text{sc}}{V_{n-1}'-V_\text{sc}}=t_{\text{eh},n}$ for $n\neq 0$ and $n\neq N^{(l)}+1$. Solving the full \cref{eq:coherent-LB-N} then implies the currents $I_1=I_1'$, $I_2=I_{N^{(l)}+2}'$, $I_3$ and voltages $V_1=V_1'$, $V_2=V_{N^{(l)}+2}'$ among the three original leads satisfy equations of the same form as \cref{eq:coherent-LB}, except that the charge transmission fractions $t_{\text{eh}}^{(u)}$ and $t_{\text{eh}}^{(l)}$ are replaced by the following effective ones:
\begin{equation}\label{eq:teh-eff}
t_{\text{eh}}^{(u)}=\prod_{n=N^{(l)}+2}^{N^{(l)}+N^{(u)}+2}t_{\text{eh},n}\ ,\quad t_{\text{eh}}^{(l)}=\prod_{n=1}^{N^{(l)}+1}t_{\text{eh},n}\ .
\end{equation}
$R_d$ and $\mathcal{P}_{\text{eh}}$ are still given by \cref{eq:Rd,eq:Peh}, but in terms of the effective charge transmission fractions in \cref{eq:teh-eff}. Besides, if the upper edge is not SC proximitized \cite{Zhao2020}, one should set $t_{\text{eh}}^{(u)}=1$ in \cref{eq:teh-eff}.

We now study the probability distribution of $\mathcal{P}_{\text{eh}}\approx t_{\text{eh}}^{(l)}$ in the form of \cref{eq:teh-eff}. Since only the lower edge is relevant, hereafter we simply denote $t_{\text{eh}}^{(l)}$ by $t_{\text{eh}}$, and $N^{(l)}$ by $N$. For $N$ metallic puddles, the problem then reduces to finding the probability distribution $P_N(t_\text{eh})$ of the variable 
\begin{equation}\label{eq:tehN}
t_\text{eh}=\prod_{n=1}^{N+1}t_{\text{eh},n}\ ,\qquad (|t_{\text{eh},n}|\le 1).
\end{equation}

The transmission fraction $t_{\text{eh},n}$ of each coherent piece of edge is assumed to obey a probability distribution function $P_0(t_{\text{eh},n})\ge0$ (normalized such that $\int_{-\infty}^\infty dy P_0(y)=1$), which is for instance given by \cref{P0} but not necessarily. Increasing $N$ by $1$, we can see a recursion relation $P_{N+1}(t_\text{eh})=\int dy dz \delta(yz-t_\text{eh}) P_{0}(y)P_N(z)$, or
\begin{equation}\label{eq:iteration}
\begin{split}
P_{N+1}(t_\text{eh})=\int_{-\infty}^\infty \frac{dy}{|y|} P_{0}(y)P_N(\frac{t_\text{eh}}{y})\ .
\end{split}
\end{equation}
Irrespective of the function $P_0(t_\text{eh})$, this equation can be formally solved by expanding $P_N(t_\text{eh})$ as a powers series of $1/|t_\text{eh}|$, which has the general solution (SM \cite{suppl})
\begin{equation}\label{eq:PN-large-t}
P_N(t_\text{eh})=\frac{\alpha_0}{|t_\text{eh}|}+\sum_{k=1}^\infty \mu_k^N \frac{\alpha_k}{|t_\text{eh}|^{k+1}}\ .
\end{equation}
Here $\mu_k=\int_{-\infty}^\infty dy |y|^k P_0(y)\ge 0$ is the $k$-th moment of probability function $P_0(y)$ for $|y|$, and $\alpha_k$ ($k\ge0$) are constants. In principle, by setting $N=0$, one can derive $\alpha_k$ by Taylor expanding $P_0(t_\text{eh})$ with respect to $1/|t_\text{eh}|$ if $P_0(t_\text{eh})$ is analytical everywhere including at $|t_\text{eh}|=\infty$. However, this is generically not true, e.g., for $P_0(t_\text{eh})$ in \cref{P0}, for which \cref{eq:PN-large-t} should be viewed as an asymptotic series for certain ranges of $t_\text{eh}$.

Nevertheless, \cref{eq:PN-large-t} is readily suggesting $P_N(t_\text{eh})$ is peaked at $t_\text{eh}=0$ when $N>0$. In particular, since $P_0(y)$ is nonzero only when $|y|\le A$ for some maximum $A\le 1$, we can estimate that $\mu_k\lesssim \frac{A^k}{k+1}<1$. This indicates that
\begin{equation}\label{eq:Pinf}
P_N(t_\text{eh})\rightarrow P_{\infty}(t_\text{eh})=\frac{\alpha_0}{|t_\text{eh}|}\ ,\quad (N\rightarrow\infty).
\end{equation}
Such a limit is ill-defined, since the total probability $\int P_\infty(t_\text{eh})d t_\text{eh}$ is divergent at $t_\text{eh}=0$ and $t_\text{eh}\rightarrow\infty$. Since physically $|t_\text{eh}|\le 1$, the divergence at $t_\text{eh}\rightarrow\infty$ is not truly problematic. In the below, we resolve the divergence at $t_\text{eh}=0$, which can hardly be seen from the $1/t_\text{eh}$ expansion.

As $t_\text{eh}\rightarrow 0$, if we assume $P_0(y)$ is around similar magnitude for $|y|\le A$ and nearly zero for $|y|>A$, it can be seen iteratively \cite{suppl} that the leading divergent term is $P_N(t_\text{eh})\propto \left[\log (A/|t_\text{eh}|)\right]^{N}$. This motivates us to write an ansatz for small $t_\text{eh}$ interpolating with \cref{eq:PN-large-t,eq:Pinf}:
\begin{equation}\label{eq:Pzero}
P_N(t_\text{eh})\simeq \frac{\alpha_0}{A} \sum_{n=0}^{N} \frac{\beta_n}{n!}\left[\log \left(\frac{A}{|t_\text{eh}|}\right)\right]^n,\quad (t_\text{eh}\rightarrow 0)
\end{equation}
with certain constants $\beta_n$. When $N\rightarrow\infty$, taking $\beta_n\rightarrow 1$ would reproduce \cref{eq:Pinf}. Note that for any finite $N$, the integral $\int P_N(t_\text{eh})dt_\text{eh}$ is not divergent at $t_\text{eh}=0$.

\emph{Monte Carlo simulation}. We further numerically evaluate the distribution $P_N(t_\text{eh})$ by sufficiently many Monte Carlo samplings. By generating each $t_{\text{eh},n}=A_0+A_1\cos\phi_n$ from uniformly random angles $\phi_n$, we calculate the distributions $P_N(t_\text{eh})$ of $t_\text{eh}$ in \cref{eq:tehN}. 

Taking $A_0=0.25$, $A_1=0.75$ as an example, which allows $t_{\text{eh},n}$ to be either positive or negative, $P_0(t_\text{eh})$ is peaked at the maximum and minimum of $t_\text{eh}$ (\cref{fig2}(a)), agreeing with \cref{P0}. For $N>0$, all $P_N(t_\text{eh})$ are peaked at $t_\text{eh}=0$ as shown in \cref{fig2}(b), as \cref{eq:Pzero} suggests. The log-log plot in \cref{fig2}(c) shows that $P_N(t_\text{eh})$ at small $t_\text{eh}$ indeed approaches the power law of \cref{eq:Pinf} as $N$ increases. SM \cite{suppl} further shows that $P_N(t_\text{eh})$ at small $t_\text{eh}$ is well-fitted by \cref{eq:Pzero} with simply $\beta_n=1$.

Since each $|t_{\text{eh},n}|\le 1$, one should expect $P_N(t_\text{eh})$ to decay faster than \cref{eq:Pinf} when $|t_\text{eh}|$ approaches its maximum. As shown in the semi-log plot \cref{fig2}(d), Monte-Carlo simulations suggest a good approximation at relatively large $|t_\text{eh}|$ is that $P_N(t_\text{eh})$ decays exponentially in $|t_\text{eh}|$, which agrees with the experimental analysis \cite{Zhao2023}. This can hardly be analyzed from \cref{eq:PN-large-t} since it is an asymptotic series. Instead, we can take a crude approximation $P_0(t_\text{eh})\sim \frac{1}{2}\delta(|t_\text{eh}|-A)$ in \cref{eq:iteration}, resembling the two peaks at $A_\pm\sim\pm A$ in \cref{P0}, where $A$ is the maximum of $|t_\text{eh}|$. \cref{eq:iteration} then implies $P_{N+1}(t_\text{eh})\sim \frac{1}{A}P_{N}(\frac{t_\text{eh}}{A})$. Considering the actual $P_{N}(t_\text{eh})$ decays to zero at $|t_\text{eh}|\sim A^{N+1}$, we estimate that
\begin{equation}\label{eq:Pdecay}
P_{N}(t_\text{eh})\sim \frac{e^{-\Gamma |t_\text{eh}|/A^{N+1}}}{A^{N+1}}\ ,\quad (t_\text{eh}\sim \mathcal{O}(A^{N+1}))
\end{equation}
with some number $\Gamma\gtrsim 1$.

In another example with $A_0=0.75$, $A_1=0.25$, one always has $t_{\text{eh},n}>0$ (which holds if the Andreev transmission $r_A$ is small) as shown in \cref{fig2}(e), and thus one finds $P_N(t_\text{eh})>0$ only when $t_\text{eh}>0$ as shown in \cref{fig2}(f). In this case, $P_N(0)=0$, while the large $t_\text{eh}$ behavior is similar to \cref{eq:PN-large-t,eq:Pdecay}.

\begin{figure}[tbp]
\centering
\includegraphics[width=3.5in]{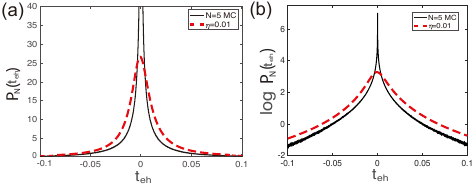}
\caption{(a) The distribution $P_N(t_\text{eh})$ for $N=5$ from Monte-Carlo (solid line) and the blurred distribution with $\eta=0.01$ (dashed line). (b) The semi-log plot of panel (a). 
}
\label{fig3}
\end{figure}

\emph{Physical considerations}. In experiment, the distribution $P_N(t_\text{eh})$ is peaked at $t_\text{eh}=0$ but may not be diverging, due to classical noises, thermal or quantum fluctuations (of pairing, etc). We include these factors as an effective Lorentzian blurring of width $\eta$, namely, $P_N(t_\text{eh})\rightarrow \int_{-\infty}^\infty \frac{\eta}{\pi(y^2+\eta^2)}dy P_N(t_\text{eh}+y)$. The dashed lines in \cref{fig3}(a) and (b) shows the blurred $P_N(t_\text{eh})$ and its logarithm, which is closer to an exponentially decaying function in \cref{eq:Pdecay} in the entire range of $t_\text{eh}$. This qualitatively agrees with the experimental data \cite{Zhao2023}.

\cref{eq:tehN} also indicates that $t_\text{eh}$ decays exponentially in the number of metallic puddles $N$ along the edge. For edge length $L_\text{edge}$, the number of magnetic vortex cores in magnetic field $B$ within a distance of SC coherence length $\xi$ to the edge is $2eB\xi L_\text{edge}/h\propto BL_\text{edge}$, which are metallic and contribute to $N$. In addition, we expect the number of metallic puddles induced by generic local SC suppressing perturbations to increase linearly in temperature $T$ and linearly in $L_\text{edge}$, as the energy density difference between metallic and SC states decreases linearly in $T$ to the leading order. Thus, $N\simeq N_0+(c_BB+c_TT)L_\text{edge}$ with $c_B>0$ and $c_T>0$ (see SM \cite{suppl}). We thus expect $t_\text{eh}$ to have a standard deviation $\sigma(t_\text{eh})\propto \sigma_0^{N+1}$, where $0<\sigma_0<1$ is the standard deviation of the function $P_0(t_\text{eh})$, or explicitly,
\begin{equation}\label{eq:deviation}
\sigma(t_\text{eh})\propto \exp\left(- \kappa_B B L_\text{edge}-\kappa_T T L_\text{edge}\right)\ ,
\end{equation}
with $\kappa_B\simeq c_B\log(\sigma_0^{-1}) >0$ and $\kappa_T\simeq c_T\log(\sigma_0^{-1}) >0$. This agrees with the scaling behavior observed in the experiment \cite{Zhao2023}. Lastly, at the critical point of losing SC entirely, we expect $N$ to effectively diverge, and thus $t_\text{eh}\rightarrow 0$.

\emph{Discussion}. Our theory shows that the probability distribution of charge transmission $t_\text{eh}$ (measurable from resistance $R_d$ in \cref{eq:Rd}) provides a signature for the coherence of superconducting chiral edge states. The distribution is peaked away from zero if coherent, and is peaked at zero if strongly decoherent. Our decoherent distribution agrees with the experiment \cite{Zhao2023}, and is distinct from the uniform distribution predicted in \cite{kurilovich2201}. While our theory assumes strong decoherence from metallic puddles (magnetic vortex cores, etc.) which serve as floating leads, we expect a similar distribution for strong decoherence of generic origins, since dephasing and energy loss can be effectively simulated by some floating leads. A future question is to extend the current study to the SC proximity \cite{jia2024superconductivity} of fractional QH and the recently realized fractional Chern insulator \cite{Cai2023,Zeng2023,Park2023,Xu2023,Lu2024,Spanton2018,Xie2021}, which is significant for probing the coherence of fractionalized edge states.

\begin{acknowledgments}
\emph{Acknowledgments}. We thank Gleb Finkelstein and Pok Man Tam for enlightening discussions. BL is supported by the National Science Foundation through Princeton University’s Materials Research Science and Engineering Center DMR-2011750, and the National Science Foundation under award DMR-2141966. Additional support is provided by the Gordon and Betty Moore Foundation through Grant GBMF8685 towards the Princeton theory program. JW is supported by the National Key Research Program of China under Grant No.~2019YFA0308404, the Natural Science Foundation of China through Grants No.~12350404 and No.~12174066, Shanghai Municipal Science and Technology Commision under Grants No.~23JC1400600 and No.~2019SHZDZX01.
\end{acknowledgments}

\bibliography{TSCedge}

\pagebreak
\widetext
\clearpage
\begin{center}
\textbf{\large Supplemental Material for ``Resistance Distribution of Decoherent Quantum Hall-Superconductor Edges"}
\end{center}
\setcounter{equation}{0}
\setcounter{figure}{0}
\setcounter{table}{0}
\setcounter{page}{1}
\makeatletter
\renewcommand{\theequation}{S\arabic{equation}}
\renewcommand{\thefigure}{S\arabic{figure}}
\renewcommand{\thesection}{\Roman{section}}

\section{I. Derivations of the coherent charge transmission for $C_{BdG}=2C=4$}\label{app:SMA}

In this section, we consider the charge transmission of a coherent QH-SC edge of a QH-SC-QH junction (as shown in main text Fig. 1(a)) with Chern number $C=2$, which is the case of the experiment \cite{Zhao2023}. The superconducting middle region has a BdG Chern number $C_\text{BdG}=2C=4$.

The propagation of the $C_\text{BdG}=4$ chiral Majorana modes at zero energy are governed by the Euler-Lagrange equation from the action in main text Eq. (1):
\begin{equation}\label{Seq:amplitudes}
v_j\partial_x \zeta_j(x) + \sum_{j'=1}^{4}M_{jj'}(x)\zeta_{j'}(x)=0\ .
\end{equation}
Here $\zeta_j(x)$ are the amplitudes of the incident chiral Bogoliubov fermion mode at zero energy, which has a creation operator $f^\dag=\sum_{j=1}^{2C}\int dx \zeta_j(x)\gamma_j$. When the velocity $v_j$ anisotropy is small, and the mass matrix $M(x)$ is random due to elastic disorders (still coherent), as shown in \cite{Wangj2018,Lian2019}, the velocity anisotropy is irrelevant, and we can approximately replace $v_j$ by their average velocity $\overline{v}$. As a result, one finds the coefficients $\bm{\zeta}(x)=(\zeta_1,\zeta_2,\zeta_3,\zeta_4)^T$ along an edge of length $L_\text{edge}$ satisfies
\begin{equation}
\bm{\zeta}(L_\text{edge})=Q\bm{\zeta}(0)\ ,\qquad Q=\mathcal{P} \text{exp}\left(-\frac{1}{\overline{v}}\int_{0}^{L_\text{edge}} M(x) dx\right)\ .
\end{equation}
As shown in \cite{Wangj2018,Lian2019}, the normal transmission coefficient subtracting the Andreev transmission coefficient is given by
\beq
t-t_A=\sum_{i,j=1}^2 (|u_j^\dagger Q u_i|^2-|u_j^T Q u_i|^2)=\sum_{i,j=1}^2 \det Q_2^{(2j-1,2j|2i-1,2i)}\ ,
\label{Q2}
\eeq 
where $u_1=(1,i,0,0)^T/\sqrt{2}$ and $u_1=(0,0,1,i)^T/\sqrt{2}$ is the annihilation operator of the chiral electron modes $c_1$ and $c_2$ written in the Majorana basis, and
\begin{equation}
Q_2^{(i,j|k,l)}=\begin{pmatrix}
Q_{ik} & Q_{il} \\
Q_{jk} & Q_{jl}
\end{pmatrix} 
\end{equation} 
is a $2\times 2$ minor matrix with $Q_{ik}$ represents the $(i,k)$ element of $Q$ (we assume $i<j$ and $k<l$). The last equality in \cref{Q2} simply follows from the definition.

In the weak disorder case, as shown in \cite{Lian2019}, we approximately have $Q=\bar{G} \Lambda \bar{G}^T$, where $\Lambda=\text{diag}(e^{i\phi_1 \sigma_y},e^{i\phi_2 \sigma_y})$ with $\sigma_y$ being the Pauli-$y$ matrix, $\bar{G}=(\mathbf{g}_1,\mathbf{g}_2, \mathbf{g}_3, \mathbf{g}_4)$ is a SO$(4)$ matrix, and the vectors $\mathbf{g}_l=(g_l^1,g_l^2,g_l^3,g_l^4)^T$ ($1\leq l\leq 4)$ form a new orthonormal Majorana basis. It is convenient to write $\bar{G}$ as a product of two unit quaternions $a_{i}+b_{i} \mathbf{i}+c_{i} \mathbf{j}+d_{i} \mathbf{k}$ ($a_i^2+b_i^2+c_i^2+d_i^2=1$ for $i=1,2$) in this case
\beq
\bar{G}=\begin{pmatrix}
a_1 & -b_1 & -c_1 & -d_1\\
b_1 & a_1 & -d_1 & c_1\\
c_1 & d_1 & a_1 & -b_1\\
d_1 & -c_1 & b_1 & a_1\\
\end{pmatrix} \cdot  \begin{pmatrix}
a_2 & -b_2 & -c_2 & -d_2\\
b_2 & a_2 & d_2 & -c_2\\
c_2 & -d_2 & a_2 & b_2\\
d_2 & c_2 & -b_2 & a_2\\
\end{pmatrix}
 \label{G2}
\eeq in matrix form. Mathematically, this is because the fact that $SO(4)\cong SU(2)\otimes SU(2)\cong SO(3)\otimes SO(3)$ (here we ignored the $\mathbb{Z}_2$ factors) as each unit quaternion represents a $SO(3)$ rotation in 3D Euclidean space. 

%
Substitute \cref{G2} into \cref{Q2}, we have

\beq
\begin{aligned}
t-t_A&=2t_\text{eh}=2((a_1^2+b_1^2-c_1^2-d_1^2)^2-4(a_1^2+b_1^2)(c_1^2+d_1^2)\cos{(\phi_1+\phi_2)})(a_2^2+b_2^2+c_2^2+d_2^2)^2\\
&=2((a_1^2+b_1^2-c_1^2-d_1^2)^2-4(a_1^2+b_1^2)(c_1^2+d_1^2)\cos{(\phi_1+\phi_2)})\\
&=2(\cos^2\psi+\cos2\theta\sin^2\psi)^2+8\sin^2\psi\sin^2\theta(\cos^2\psi+\cos^2\theta\sin^2\psi)\cos(\phi_1+\phi_2)\\
&=2A_0+2A_1\cos(\phi_1+\phi_2),
\end{aligned}
\eeq 
where 
\begin{equation}
A_0=(\cos^2\psi+\cos2\theta\sin^2\psi)^2\ ,\qquad  A_1=4\sin^2\psi\sin^2\theta(\cos^2\psi+\cos^2\theta\sin^2\psi)\ ,
\end{equation}
and the angles come from the parametrization of the first unit quaternion as $a_1=\sin\psi\sin\theta\sin\varphi$, $b_1=\sin\psi\sin\theta\cos\varphi$, $c_1=\sin\psi\cos\theta$, $d_1=\cos\psi$, where $\psi\in[0,\pi]$, $\theta\in[0,\pi]$, and $\varphi\in[0,2\pi]$. Interestingly, the resulting $t-t_A=2t_\text{eh}$ does not depend on the second quaternion. This is because the second quaternion corresponds to the SU(2) rotation in the electron flavor basis, which does not contribute to the electron-hole conversion, and thus does not affect $t_\text{eh}$. Redefining $\phi=\phi_1+\phi_2$, we obtain the charge transmission in the main text Eq. (3), namely,
\begin{equation}
t_{\text{eh}}= A_0+A_1\cos\phi\ .
\end{equation}
The probability distribution $P_0(t_{\text{eh}})$ of the charge transmission $t_{\text{eh}}$ can be calculated as
\beq
\begin{aligned}
P_0(t_\text{eh})&=\frac{\int_0^{2\pi} \delta(t_\text{eh}-A_0-A_1\cos\phi) d\phi}{\int_0^{2\pi} d\phi}=\frac{1}{\pi}\int_0^\pi \delta(t_\text{eh}-A_0-A_1\cos\phi)d\phi =\frac{1}{\pi A_1 \sin\phi}\Big|_{A_0+A_1\cos(\phi)=t_\text{eh}}\\
&=\begin{cases}
&\frac{1}{\pi \sqrt{A_1^2-(t_{\text{eh}}-A_0)^2}}\ , \ (t_{\text{eh}}\in [A_-,A_+])\\
&0\ ,\qquad \qquad\qquad\ \    (t_{\text{eh}}\notin [A_-,A_+])
\end{cases}
\end{aligned}
\eeq 
where $A_\pm=A_0\pm |A_1|$. This gives the probability distribution in main text Eq. (7).

\section{II. Formal derivation of the generic distribution function}

In this section, we derive in more details the general solution of the probability distribution $P_{N}(t_\text{eh})$ in different limits. We first recall the iteration equation $P_{N}(t_\text{eh})$ satisfies:
\begin{equation}
P_{N+1}(t_\text{eh})=\int_{-\infty}^\infty dy dz \delta(yz-t_\text{eh}) P_{0}(y)P_N(z)=\int_{-\infty}^\infty \frac{dy}{|y|} P_{0}(y)P_N(\frac{t_\text{eh}}{y})= \int_{-\infty}^\infty \frac{dz}{|z|} P_{0}(\frac{t_\text{eh}}{z})P_N(z)\ .
\end{equation}
Assume $P_{N}(t_\text{eh})$ has an expansion as powers of $1/|t_\text{eh}|$:
\begin{equation}
P_{N}(t_\text{eh})=\sum_{k=0}^\infty \frac{J_k}{|t_\text{eh}|^{k+1}}\ ,
\end{equation}
where $J_k$ are coefficients to be determined. The series starts from power $1/|t_\text{eh}|$ since a physical probability distribution must vanish at $|t_\text{eh}|=\infty$. The iteration equation then yields:
\begin{equation}
P_{N+1}(t_\text{eh})=\int_{-\infty}^\infty \frac{dy}{|y|} P_{0}(y)\sum_{k=0}^\infty \frac{J_k|y|^{k+1}}{|t_\text{eh}|^{k+1}} =\sum_{k=0}^\infty \frac{J_k}{|t_\text{eh}|^{k+1}} \int_{-\infty}^\infty  dy |y|^k P_{0}(y)= \sum_{k=0}^\infty \mu_k\frac{J_k}{|t_\text{eh}|^{k+1}}\ ,
\end{equation}
where we have defined the $k$-th moment
\begin{equation}
\mu_k=\int_{-\infty}^\infty dy |y|^k P_0(y)\ge 0\ .
\end{equation}
By iteration, this implies a generic solution that $J_k=\alpha_k \mu_k^N$, namely,
\begin{equation}\label{Seq:PN-large-t}
P_N(t_\text{eh})=\frac{\alpha_0}{|t_\text{eh}|}+\sum_{k=1}^\infty \mu_k^N \frac{\alpha_k}{|t_\text{eh}|^{k+1}}\ ,
\end{equation}
where $\alpha_k$ are constants depending on the function $P_0(t_\text{eh})$.

We further examine the small $t_\text{eh}\rightarrow 0$ limit. In a crude approximation, assume $P_0(t_\text{eh})\simeq \frac{1}{2A}\Theta(A-|t_\text{eh}|)$ is approximately uniform within $|t_\text{eh}|<A$ and zero otherwise. The iteration equation then indicates
\begin{equation}
P_{N+1}(t_\text{eh})= \int_{-\infty}^\infty \frac{dz}{|z|} P_{0}(\frac{t_\text{eh}}{z})P_N(z)\simeq\frac{1}{A}\int_{\frac{|t_\text{eh}|}{A}}^\infty d\left(\log |z|\right) P_N(z)\ .
\end{equation}
Starting from $N=0$, and since $P_0(z)$ is approximately constant as $z\rightarrow 0$, we find the leading divergent term
\begin{equation}
P_1(t_\text{eh})\propto \log (A/|t_\text{eh}|)\ ,\quad P_2(t_\text{eh})\propto \left[\log (A/|t_\text{eh}|)\right]^{2}\ ,\quad \cdots \quad P_N(t_\text{eh})\propto \left[\log (A/|t_\text{eh}|)\right]^{N}\ .
\end{equation}
This leads to the generic form in the $t_\text{eh}\rightarrow 0$ limit conjectured in our main text.

\section{III. Monte-Carlo simulation}\label{app:MC}
We describe in more details our Monte-Carlo numerical simulations for the probability distribution $P_N(t_\text{eh})$. The target charge transmission $t_\text{eh}=\prod_{n=1}^{N+1}t_{\text{eh},n}$ takes a product form. We randomly generate each $t_{\text{eh},n}=A_0+A_1\cos\phi_n$ from uniformly random angles $\phi_n\in[0,2\pi)$, and calculate $t_\text{eh}$. By sampling up to $10^8$ times, we calculate its distribution $P_N(t_\text{eh})$ for a given $N$.

\begin{figure}[htbp]
\centering
\includegraphics[width=3.5in]{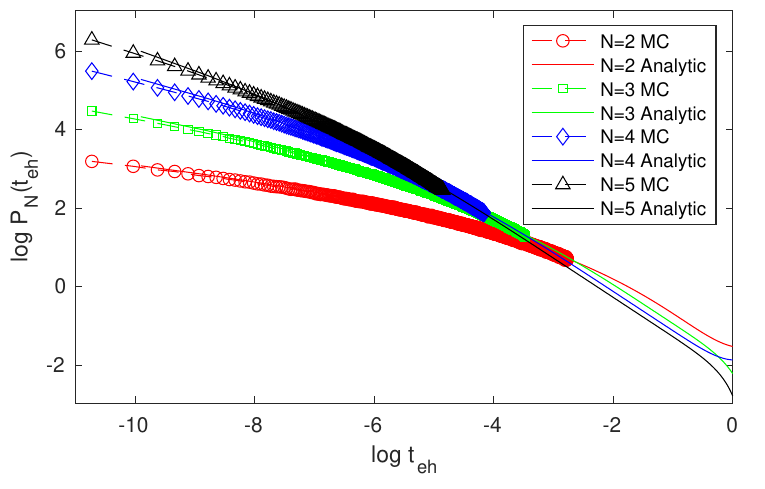}
\caption{Fitting the Monte-Carlo (MC) distribution $P_N(t_\text{eh})$ data (the dots) by the analytical ansatz \cref{Seq:Pzero} (the solid lines), shown in the log-log (the natural log) plot. The number of metallic puddles $N$ is listed in the legend.}
\label{figS}
\end{figure}

Moreover, to test whether the main text Eq. (14) gives a good ansatz for the distribution $P_N(t_\text{eh})$ at small $t_\text{eh}\rightarrow 0$, we fit the Monte Carlo data by setting $\beta_n=1$ (note that we have argued that $\beta_n\rightarrow 1$ as $N\rightarrow\infty$), and treat $\alpha_0$ and $A$ as the only two fitting parameters. Namely, we take the ansatz

\begin{equation}\label{Seq:Pzero}
P_N(t_\text{eh})\simeq \frac{\alpha_0}{A} \sum_{n=0}^{N} \frac{1}{n!}\left[\log \left(\frac{A}{|t_\text{eh}|}\right)\right]^n\ .
\end{equation}
The fitting results for small $t_\text{eh}$ is shown in the log-log plot in \cref{figS}, which shows nice agreement for different values of $N$.

\section{IV. Estimation of the number of metallic puddles}

We briefly estimate the number of metallic puddles along a QH-SC edge of length $L_\text{edge}$. In a magnetic field $B$, a magnetic vortex core which has distance within a SC coherence length $\xi$ to the edge would be a metallic puddle encountered by the chiral edge state. Since each vortex carries a magnetic flux $h/2e$, the number of such vortex cores can be estimated from the total magnetic flux within an area $\xi L_\text{edge}$ around the edge:
\begin{equation}
N_\text{mag}\simeq\frac{B \xi L_\text{edge}}{h/2e}=\frac{2e B \xi L_\text{edge}}{h}\ .
\end{equation}
In addition, metallic puddles may arise due to other SC suppressing perturbations, such as local repulsive interactions. Within the Ginzburg Landau theory, in the absence of such perturbations, the metal phase is energetically higher than the SC phase by an energy density difference $\Delta u(T)\simeq a(T-T_c)^2\approx aT_c^2-2aT_c T$ for some $a>0$, which linearly decreases with respect to temperature $T$. Assume a local perturbation lowers the energy density of the metal phase relative to the SC phase by $w$, the metal phase will be favored if $\Delta u(T)<w$. Assuming such perturbation energy densities $w$ have a sufficiently uniform distribution function $p(w)\approx \frac{1}{W_0}$ within a range of $w\in [0,W_0]$, the number of metallic puddles from such origins would be
\begin{equation}
N_\text{th}\simeq \frac{L_\text{edge}}{l_\text{pert}} \int_{\Delta u(T)}^{W_0} p(w)\approx \frac{L_\text{edge}}{l_\text{pert}}\left(1-\frac{aT_c^2}{W_0}+\frac{2aT_c T}{W_0}\right)\ ,
\end{equation}
where $l_\text{pert}$ is the typical size of such a perturbed region which depends on the system details. Thus, the total number of metallic puddles has the form
\begin{equation}
N\simeq N_\text{mag}+N_\text{th}\simeq N_0+(c_BB+c_TT)L_\text{edge}\ ,
\end{equation}
with $c_B\simeq \frac{2e \xi}{h}$ and $c_T\simeq \frac{2aT_c }{W_0 l_\text{pert}}$ being positive numbers. This then leads to the exponential decay of standard deviation in $B$ and $T$ in main text Eq. (16).

As an estimation, for $\xi\sim 10$~nm (short since the proximity induced SC in 2D is type II), one has $c_B\sim 30~$T$^{-1}\mu$m$^{-1}$, and then $\kappa_B\simeq c_B\log(\sigma_0^{-1})$ in the main text Eq. (16). In the experiment \cite{Zhao2023}, $\kappa_B\sim 1~$T$^{-1}\mu$m$^{-1}$, suggesting $\log(\sigma_0^{-1})\sim 10^{-1}$, which would be consistent if the coherent standard deviation $\sigma_0<1$ is close to $1$.

\end{document}